\documentclass[12pt,preprint]{aastex}

\usepackage{natbib}
\usepackage{graphicx}

\begin{document}

\newcommand{\lya}{Lyman-$\alpha$}
\newcommand{\eqw}{\hbox{EW}}
\def\erg{\hbox{erg}}
\def\cm{\hbox{cm}}
\def\sec{\hbox{s}}
\def\f17{f_{17}}
\def\Mpc{\hbox{Mpc}}
\def\Gpc{\hbox{Gpc}}
\def\nm{\hbox{nm}}
\def\km{\hbox{km}}
\def\kms{\hbox{km s$^{-1}$}}
\def\yr{\hbox{yr}}
\def\Myr{\hbox{Myr}}
\def\Gyr{\hbox{Gyr}}
\def\deg{\hbox{deg}}
\def\arcsec{\hbox{arcsec}}
\def\microJy{\mu\hbox{Jy}}
\def\zre{z_r}
\def\fesc{f_{\rm esc}}

\def\ergcm2s{\ifmmode {\rm\,erg\,cm^{-2}\,s^{-1}}\else
                ${\rm\,ergs\,cm^{-2}\,s^{-1}}$\fi}
\def\ergsec{\ifmmode {\rm\,erg\,s^{-1}}\else
                ${\rm\,ergs\,s^{-1}}$\fi}
\def\kmsMpc{\ifmmode {\rm\,km\,s^{-1}\,Mpc^{-1}}\else
                ${\rm\,km\,s^{-1}\,Mpc^{-1}}$\fi}
\def\cMpc{\ifmmode {cMpc}\else
                ${cMpc}$\fi}
\def\kpc{{\rm kpc}}
\def\oii{[O{\sc II}] $\lambda$3727}
\def\oiipair{[O{\sc II}] $\lambda \lambda$3726,3729}
\def\oiii{[O{\sc III}] $\lambda$5007}
\def\oiiipair{[O{\sc III}]$\lambda \lambda$4959,5007}
\def\taulya{\tau_{Ly\alpha}}
\def\taubar{\bar{\tau}_{Ly\alpha}}
\def\llya{L_{Ly\alpha}}
\def\ldlya{{\cal L}_{Ly\alpha}}
\def\nbar{\bar{n}}
\def\Msun{M_\odot} 
\def\mdyn{M_{dyn}}
\def\vmax{v_{max}}
\def\sqamin{\Box'}
\def\l43{L_{43}}
\def\ls{{\cal L}_{sym}}
\def\snrat{\ifmmode {\cal S / N}\else
                   ${\cal S / N}$\fi}
\def\siglos{\sigma_{\hbox{los}}}
\def\asf{\alpha_{SF}}
\def\bsf{\beta_{SF}}
\def\SFR{\hbox{SFR}}
\def\rhoeff{\bar{\rho_e}}
\def\MsunYr{M_\odot\,\hbox{yr}^{-1}}
\title{
The Dynamical Masses, Densities, and Star Formation
Scaling Relations of Lyman-$\alpha$ Galaxies
}

\author{
James E. Rhoads\altaffilmark{1},
Sangeeta Malhotra\altaffilmark{1},
Steven L. Finkelstein\altaffilmark{2},
Johan P. U. Fynbo\altaffilmark{3},
Emily M. McLinden\altaffilmark{1,2},
Mark L. A. Richardson\altaffilmark{1},
Vithal S. Tilvi\altaffilmark{4}
}

\begin{abstract}
  We present the first dynamical mass measurements for \lya\ galaxies
  at high redshift, based on velocity dispersion measurements from
  rest-frame optical emission lines and size measurements from HST
  imaging, for a sample of nine galaxies drawn from four surveys.  
  These measurements enable us to study the nature of 
  \lya\ galaxies in the context of galaxy scaling relations. 
  The resulting dynamical masses range from $10^9$ to
  $10^{10} \Msun$.  We also fit stellar population models to our
  sample, and use them to 
  plot the \lya\ sample on a stellar mass vs. line width relation.
  Overall, the \lya\ galaxies follow well the scaling relation
  established by observing star forming galaxies at lower redshift
  (and without regard for \lya\ emission), though
  in  $\sim 1/3$ of the  \lya\ galaxies, lower-mass fits are also
  acceptable.  In all cases, the {\it dynamical} masses agree with
  established {\it stellar} mass-linewidth relation.  
  Using the dynamical masses as an upper limit on gas mass, we show 
  that \lya\ galaxies resemble starbursts (rather than ``normal'' galaxies)
  in the relation between gas mass surface density
  and star formation activity, in spite of relatively modest
  star formation rates.  Finally, we examine the
  mass densities of these galaxies, and show that their future evolution
  likely requires dissipational (``wet'') merging.  In short, we find
  that \lya\ galaxies are low mass cousins of larger starbursts.
\end{abstract}
\altaffiltext{1}{School of Earth and Space Exploration,
Arizona State University, Tempe, AZ 85287; email James.Rhoads@asu.edu}
\altaffiltext{2}{
Department of Astronomy,
University of Texas at Austin, 1 University Station C1400,
Austin, TX 78712, USA
}
\altaffiltext{3}{DARK Cosmology Centre, Niels Bohr Institute, 
University of Copenhagen, 
Juliane Maries Vej 30, 2100 Copenhagen \O, Denmark}
\altaffiltext{4}{
George P. and Cynthia W. Mitchell Institute for Fundamental Physics and 
Astronomy, Department of Physics, Texas A\&M University, College Station, 
TX 77843}

% ApJ format

\section{Introduction}
Lyman $\alpha$ line emission is an increasingly important tool for
identifying actively star forming galaxies in the distant universe.  
First proposed as a potential signpost for primitive galaxies \citep{PP67},
\lya\ emission has now been used to identify thousands 
of galaxies at redshifts $2 < z < 7$, along with smaller samples
at $z<2$ and several candidates at $z>7$.  

While no galaxy yet identified by any means is demonstrably
primordial, \lya\ selected samples do have several properties
suggestive of youth.  Their starlight is dominated by young
populations with characteristically low stellar masses
\citep{Pirzkal07,Finkelstein07} and small sizes \citep{Bond09,Malhotra12}.
Yet, the correlation properties of these objects suggest that they are
associated with moderately large halos (mass $\sim 10^{11} \Msun$;
\citet{kovac07,guaita10}).  Combining these results suggests that a \lya\ galaxy
contains only a small fraction of the baryons that should be associated
with its host dark matter halo.  It would be interesting to know whether the
``missing'' baryons are present, either as old stellar populations, 
or in the interstellar medium of the \lya\ galaxy.

Dynamical mass estimates for \lya\ galaxies could potentially address
this question, providing a standard for comparison with both the
stellar masses and the dark halo masses.  However, such dynamical mass
estimates require an accurate measurement of the galaxy's velocity
dispersion.  The easiest approach to kinematics would be to use the
\lya\ line width, which is measured for most spectroscopically
confirmed \lya\ galaxies.  Unfortunately, this does not lead to
useful velocity dispersion information: The \lya\ line profile can be
dramatically affected by the interplay of resonant scattering and gas
kinematics in the emitting galaxy.  

We therefore turn in this paper to studying \lya\ galaxy kinematics
using the strong rest-frame optical emission lines 
of [OIII] $\lambda 5007$\AA\ and H$\alpha$ $\lambda6561$\AA.  
We build on the first detections of such lines 
in \lya\ selected galaxies at redshifts $2.2 \la z \la 3.1$
\citep{McLinden11,Finkelstein11a,Hashimoto12}.  
In section~\ref{sec:samp_dat}, we describe the sample and the key data.
In section~\ref{sec:dynmass1} we estimate dynamical masses
based on the observed line widths.
In section~\ref{sec:popsynth}, we analyze the stellar populations 
and dust reddening in
these galaxies, using deep archival photometry.
In section~\ref{sec:tf}, we combine the
rest-optical line width measurements with the stellar masses from population
synthesis modelling to compare these \lya\ galaxies with expectations
from the stellar mass Tully-Fisher relation. 
In section~\ref{sec:sfr_scale}, we examine the relation between 
gas mass surface density and star formation surface density
to show \lya\ galaxies lie on the same sequence as starburst galaxies.
Finally, in section~\ref{sec:density},
we explore the mass densities of the sample and the implications
for the future evolution of \lya\ galaxies.

Throughout the paper, we adopt a $\Lambda$-CDM ``concordance
cosmology''  with $\Omega_M = 0.27$, $\Omega_\Lambda = 0.73$, and
$H_0 = 71 \kmsMpc$.

\section{Description of Sample}
\label{sec:samp_dat}
Our sample consists of \lya\ emitting galaxies that were selected from
four surveys.  All are selected directly by the presence of a strong
\lya\ emission line in the survey data.  For the analysis in this
paper, we make use of (a) velocity dispersions $\sigma$, derived from
spectra of rest frame optical emission lines; (b) sizes, as measured
by the half-light radius $r_e$ in broad-band optical Hubble Space Telescope
images; 
(c) stellar masses and dust extinctions, derived from spectral
energy distribution (SED) fits; and 
(d) star formation rates (SFR), determined from H$\alpha$ line flux
measurements where available (and from spectral energy distribution
fitting otherwise), and corrected for dust using the results of
the SED fits.
In this section we summarize the
sources for various galaxies in the sample, along with the measurements
of $\sigma$, $r_e$, and SFR.  The stellar mass derivations are discussed
later, in section~\ref{sec:popsynth}.  The properties
of the sample are also summarized in tables 1--3.

\subsection{Bok telescope $z=3.1$ survey objects:}
First, we select two \lya\ galaxies from a 5020\AA\ narrowband survey
using the Steward Observatory's 90 inch Bok Telescope on Kitt Peak,
Arizona. The objects we study here were spectroscopically confirmed as
\lya\ emitters at $z\approx 3.12$ using the 6.5m MMT on Mt.\ Hopkins,
Arizona.  This survey was introduced in \citet{McLinden11}, and
further details will be presented in a forthcoming paper
\citep{McLinden12}.  We have followed up galaxies from this sample
with three near-infrared spectrographs: LUCIFER, on the Large Binocular
Telescope \citep{McLinden11}; NIFS, on Gemini North
\citep{Richardson12}; and NIRSPEC, on the Keck II telescope
\citep{McLinden12}.  The objects studied here are LAE40844
and LAE27878 (following the numbering in \citet{McLinden11}).

The primary source of line widths for these objects is our Gemini
NIFS data \citep{Richardson12}, 
since our Keck observations of 40844 were primarily aimed at the 
(still undetected) \oiipair\ lines, while the kinematic line width is
not well resolved in the LUCIFER observations.
These Gemini observations yielded line widths of $\Delta V_{\rm FWHM} =
13.8$\AA\ for object 40844 and 8.6\AA\ for object 27878.  The
instrumental resolution was about 5\AA.  Subtracting this instrumental
resolution in quadrature from the measured line widths yields 13.0\AA\
and 7.0\AA\ respectively.  The corresponding line-of-sight velocity 
dispersions become $\siglos = 80\kms$ and $\siglos = 43 \kms$ respectively.

The half-light radii for these objects are 1.1 kpc for LAE40844
and 1.3 kpc for LAE27878 \citep{Malhotra12}. 
We estimated their star formation rates from SED fitting 
(section~\ref{sec:popsynth}), since their H$\alpha$ line falls
at $2.70 \mu m$ and cannot be easily observed.  We obtained  
$120 \Msun \yr^{-1}$ for LAE40844, and 
$34 \Msun \yr^{-1}$ for LAE27878.  We estimate the uncertainties
in these star formation rates from the range of SFR in models yielding 
acceptable fits to the data (see section~\ref{sec:popsynth}).

\subsection{HETDEX Pilot Survey objects:}
Second, we use two \lya\ galaxies selected using a blind spectroscopic
search with an integral field spectrograph as part of the 
HETDEX Pilot Survey. 
These are objects HPS194 ($z=2.287$) and HPS256 ($z=2.491$).  
For these, we base our
kinematic line widths on Keck+NIRSPEC H$\alpha$ line measurements from 
\citet{Finkelstein11a,Song13}.  These used low-resolution
mode with the NSPEC-7 blocking filter.  
The measured line widths were $18.1 \pm 0.6$\AA\ (FWHM) for HPS 194, 
and $19.4\pm 0.9$\AA\ (FWHM) for HPS 256.   The instrumental resolving
power is $R \sim 1500$ with the $0.76''$ slit, corresponding to 14--15\AA\ 
FWHM, and direct measurements of sky lines in the two spectra yield
resolutions of $14.4$\AA\ and $14.1$\AA\ (FWHM) respectively.  
Subtracting these in quadrature yields line widths of 
$11.0$\AA\ and $13.3$\AA, respectively.
The corresponding line-of-sight velocity dispersions become
$\siglos = 65\kms$ and $\siglos = 74 \kms$ respectively.
The half-light radii, as reported in the COSMOS ACS i-band
catalog \citep{Leauthaud07}, are $1.5\kpc$ for HPS194, 
and $1.1 \kpc$ for HPS256. 
(HPS 194 in fact corresponds to a pair of continuum sources separated
by $\sim 0.5''$ in the COSMOS survey's HST+ACS images.  We assume that
the strong \lya\ and H$\alpha$ emission come from the brighter and
more compact source.  The other source has a $2.4 \kpc$ half-light
radius, which would raise our dynamical mass for this object by about
60\%.)

The H$\alpha$ line
fluxes \citep{Song13} yield star formation rates
of $17 \MsunYr$ and $20 \MsunYr$, respectively, before dust
correction.  This is 
based on the conversion $\SFR = 4.6\times 10^{-42} (L_{H\alpha} / \ergsec)
\MsunYr$, which is appropriate for a Chabrier (2003) initial
mass function [IMF] \citep{Twite12}.
The inferred SFR would be $1.2\times$ greater for a Kroupa (2003) IMF,
and $1.8\times$ greater for the IMF assumed in Kennicutt (1998).
We correct these star formation rates for dust extinction within
the emitting galaxy.  The uncertainty in the star formation rate
is dominated by the uncertainty in this dust correction, which we
estimate by considering the full range of extinction among models
with $\chi^2 \le \chi^2_{min} + 1$, and further allowing the 
possibility that the extinction of H$\alpha$ could exceed the
extinction of continuum starlight by up to a factor of 2.

\subsection{Subaru NB survey objects:}
Third, we use four \lya\ galaxies from a narrowband survey described
by \citet{Hashimoto12} and \citet{Nakajima12b}.
These objects are drawn from two fields--- COSMOS (objects 
COSMOS13636 and COSMOS30679)  and the Chandra Deep
Field South (objects CDFS3865 and CDFS6482).
We obtained line widths for all sources by measuring the plotted FWHM
of emission line profiles in figures~1, 2, and 5 of \citet{Hashimoto12}. 
In general, the lines in this sample appear marginally resolved.
In our analysis we regard the line width measurements as upper limits
where appropriate.

The COSMOS field spectra were from Keck + NIRSPEC spectroscopy.
Here we obtained directly measured FWHM of $16.6$ and $16.5$\AA\ for
COSMOS13636 and COSMOS30679, respectively.  With a resolving
power of $R=1500$, the instrumental resolution corresponds to $13.8$\AA,
which we subtract in quadrature to yield nominal velocity dispersions
$\sigma = 57\kms$ and $\sigma = 55 \kms$, respectively.  In each case,
the measurement remains consistent with a fairly broad range,
$0 < \sigma \la 96 \kms$.  

For these COSMOS field sources, we base the half light radii on
the COSMOS ACS i-band catalog, obtaining $0.79\kpc$ and $1.82\kpc$
respectively.
Finally, their star formation rates as inferred from their H$\alpha$ 
fluxes are $8.7 \MsunYr$ and $10.5 \MsunYr$ respectively.

The CDFS field source spectra \citep{Hashimoto12} 
were obtained with Magellan + MMIRS, and
have a somewhat lower resolving power ($R\approx 1120$, corresponding
to 18\AA).  The measured FWHM are 23\AA\ for CDFS3865 and 20\AA\ for
CDFS6482.  Subtracting the instrumental resolution in quadrature 
yields velocity dispersions of $105\kms$ and $42\kms$, respectively,
where again there is considerable uncertainty for the narrower,
semi-resolved line (consistent with $0<\sigma<71 \kms$).
The star formation rates for these sources, based on their H$\alpha$ 
line fluxes, are $125 \MsunYr$ and $31 \MsunYr$ respectively.

For the physical sizes of the CDFS sources, we downloaded archival
HST imaging from the GEMS survey \citep{Rix04} 
and measured the half light radii.
We used the SExtractor \citep{Bertin96} half light radius, and as a consistency
check also measured the fluxes in a series of circular apertures and 
interpolated the resulting photometric growth curve. Both methods gave
consistent answers, with half light radii of $0.96\kpc$ for
CDFS3865, and $1.78\kpc$ for CDFS6482.

\subsection{ESO $z=2.25$ survey object:}
Finally, we use one \lya\ galaxy, LAE-COSMOS-47, from a
narrowband-selected $z=2.25$ COSMOS field sample obtained 
by \citet{Nilsson11} using the ESO 2.2m telescope,
with followup observations obtained by Fynbo and
collaborators using the X-Shooter spectrograph
\citep{Vernet11} on the VLT.
The X-Shooter spectrum provided 
a well constrained velocity dispersion measurement of $30\kms$.
The star formation rate, derived from the H$\alpha$ line flux, is
$33\MsunYr$.  The half-light radius, from the COSMOS ACS catalog
\citep{Leauthaud07}, is $1.14\kpc$.

\section{Dynamical Mass Estimates}
\label{sec:dynmass1}
The measured velocity dispersions $\sigma$ from the rest-optical
emission lines are a good estimate of the
total luminosity-weighted kinematics of the gas.  The galaxies we are
studying are spatially unresolved even in $\sim 0.5''$ seeing, meaning that
the ground-based spectra we use effectively sample the integrated
light, with no important dependence on slit width.

The precise conversion from velocity width to mass will depend on the
kinematic structure of these galaxies.  For a pure rotation-supported
model with a flat rotation curve, we expect $\sigma^2 = \sin^2{i} \,
v_c^2 / 2$, where $i$ is the inclination angle of the disk (with
$i=90^\circ$ corresponding to an edge-on system).

The simplest dynamical mass estimate from these measurements, which is
a lower bound to the true gravitating mass, is $\mdyn \ge v_c^2 r /
G$, where $r$ is the maximum radius at which we observe light from the
galaxy.  A more practical choice of radius is the effective 
radius $r_e$, defined as the radius that encloses half of the galaxy's
light in projection.  If we presume that the half-light radius is also
the half-mass radius, our revised estimate of the dynamical mass
becomes $\mdyn \approx 2 v_c(r_e)^2 r_e / G 
\approx 4 \sigma^2 r_e / (G \sin^2{i})
\ga 4 \sigma^2 r_e / G$.
The resulting mass estimates range from $10^9$ to $10^{10} \Msun$,
and are summarized in table~\ref{tab:masses}.

Several circumstances could affect our estimated mass.  A rotating
disk with $i<90^\circ$ would reduce the measured $\sigma$ (relative to
the edge-on case).  Also, in our spatially unresolved spectroscopy,
the observed $\sigma^2$ reflects the luminosity-weighted average
kinematics.  If a significant fraction of the galaxy's light is
emitted from regions where the local circular speed $v_c(r)$ is below
the maximum circular speed ($\vmax$), we should expect the weighted
average $\sigma^2$ to underestimate $\vmax$ and hence the mass.  The
precise magnitude of this effect depends on the galaxy's light 
profile and rotation curve.
Also, like any dynamical mass based on luminous tracers, our estimate
is insensitive to mass located outside the luminous matter distribution 
of the galaxies.  If the galaxies are embedded in extended dark matter halos,
the total mass of the halo could be many times the mass estimates
derived from the observed $r_e$ and $\sigma^2$.

Turbulence in the galaxy's gas would contribute to
the measured $\sigma$, though in virial equilibrium, that turbulence would
constitute a source of pressure support and the
ordered rotation of the galaxy would be correspondingly reduced.  Most
interestingly, if the galaxy is not in equilibrium at all, our assumed
relations between kinematics and mass could be substantially wrong.
Our mass estimate implicitly assumes that the virial theorem is fulfilled,
that is, that the kinetic energy $K$ and potential energy $U$ of the
galaxy are related by $K = -U/2$. 
On the other hand, in a cold accretion scenario, new material falling into
a galaxy for the first time should have $K = -U$, i.e., the motions are
faster for the same gravitating mass under these conditions, and the mass
inferred from gas motions would be correspondingly over-estimated by a
factor up to $\sim 2$.

Given these uncertainties, it is best to regard our direct dynamical
mass estimates as approximate numbers, good to a factor of perhaps 2
when regarded as lower bounds to the true dynamical mass.  Other 
dynamical mass estimates in the literature consider a more general
scaling coefficient so that $\mdyn = \beta \sigma^2 r_e / G$
(see \citet{Toft12} and references therein, esp. \citet{Jorgensen96}
and \citet{Cappellari06}).  These works favor $\beta\approx 5$ for
early-type galaxies with Sersic index $n\approx 4$, and find that
despite theoretical expectations for some increase of $\beta$ with
decreasing Sersic $n$, the observational evidence favors $\beta \approx 5$
for a wide range of $n$.  Thus, the simple arguments that led us to use
$\beta=4$ likely come fairly close to the correct dynamical masses.

A complementary approach to interpreting the kinematic data on these
galaxies is to use their linewidths to place them on some form of the
Tully-Fisher relation, and so to compare them on an equal footing to
other galaxy populations.  Such an approch avoids the difficulties
associated with identifying the right radius to use in estimators of
the form $M\sim v^2 R / G$.  For high redshift galaxy populations, a
small scatter with weak redshift evolution has been demonstrated for
the stellar mass Tully-Fisher relation, and we place our galaxies on
such a relation in section~\ref{sec:tf} below.  To do so, we first
need their stellar masses.

\section{Population Synthesis Modeling}
\label{sec:popsynth}
All of the galaxies we study have extensive multiband photometry
in the literature, generally including multiband optical data, some
deep ground-based near-infrared photometry, and Spitzer IRAC
\citep{Fazio04} observations that are deep enough to be constraining in at 
least the 3.6$\mu m$ channel.  

We have used these data to derive stellar mass estimates 
for the full sample. 
Stellar mass estimates are also available in
the published literature for many of these galaxies
\citep{Finkelstein11a,Hashimoto12,Nakajima12b,McLinden12}.
While these are mostly consistent with our estimates where
samples overlap, we opted to fit the entire sample using
a single procedure to avoid potential difficulties comparing
masses derived using different methodologies.

\subsection{General comments on SED fitting:} 
(1) The strong \lya\ emission in these objects requires the presence
  of a young stellar population, whose ultraviolet light ionizes
  interstellar hydrogen that then recombines to produce the observed
  \lya\ radiation.
(2) The amount of dust in the fitting is essentially determined by
  the UV spectral slope (cf. \citet{Meurer97,Hathi08,Finkelstein11a}).
(3) In some objects, no stars older than $10^7$ years are {\it required} 
  to explain the observed light.
(4) A considerable mass in old stars is {\it permitted.}  
(5) The ionizing photon production for the best-fit stellar
  populations can be converted to a \lya\ luminosity by assuming that
  $2/3$ of the ionizing photons are ultimately converted to \lya\
  radiation (as expected under Case B recombination with a negligible
  escape fraction for ionizing photons).  Combining this line
  luminosity estimate with the directly measured \lya\ line flux gives
  an estimate of the escape fraction for the \lya\ photons.  These
  escape fractions are sensitive to the details of the star formation
  history over the last $\sim 6$ Myr.  We can say with reasonable
  confidence that the resulting escape fraction is of order half for
  the most plausible models, and that \lya\ escape fractions below 20--30\%
  are ruled out unless we change the stellar population in some way
  that dramatically increases the ionizing photon production.
(6) The estimated stellar population ages and masses would increase
  dramatically if we did not account for the \oiiipair\ line fluxes.
  In this case, the stellar population fitting code attempts to interpret
  the red H$-$K color and the bright flux in the rest-frame 5000\AA\ range 
  as due to older stellar populations.  This effect can exceed an order of 
  magnitude in both age and mass when the filter containing the \oiiipair\ 
  lines is the reddest filter considered, while it is usually smaller
  when an additional filter redward of the 4000\AA\ break is included
  (meaning, in our case, the IRAC photometry).

\subsection{Starburst99 modeling:} 
We measured stellar masses for our sample using the
Starburst99 population synthesis code \citep{Leitherer99}.  
For each galaxy we started with publically available broad band 
photometry.  For the seven objects in the COSMOS field, we
used COSMOS project broad band photometry \citep{Capak07,McCracken10},
either directly from the COSMOS archive, or as quoted in 
the papers defining the samples \citep{Finkelstein11a,
Hashimoto12,Nakajima12b}.  For the two CDFS objects, we used
MUSYC survey photometry \citep{Gawiser06} (again as quoted
in \citet{Hashimoto12,Nakajima12b}). In all cases, our final model fitting
used at least 9 photometric bands spanning 
at least the wavelength range $0.4 \mu m < \lambda < 3.6 \mu m$.

To interpret the photometry in terms of stellar population parameters, 
we first ran Starburst99 to generate a grid of model spectra for
stellar populations at a range of ages, from $2\times 10^6$ years
(which is so young that no star has yet left the main sequence), up
to $2\times 10^9$ years (which is the age of the universe at $z\approx
3.1$), and using a Kroupa (2003) IMF.  
We then assumed a star formation history (as described below) 
and generated a model
spectrum by a linear combination of the spectra for particular age
steps from the Starburst99 output.

We account for the mean opacity of the intergalactic medium using the 
prescription of \citet{Madau95}.  We do not treat the variance in the
IGM opacity in the present work. (Doing so would
effectively add uncertainty to the expected fluxes in the 
$u^*$, B$_j$, and $g$ bands, and so would improve the model $\chi^2$,
but would not likely change the best fit model parameters much.)

We model dust in our sample galaxies using the Small Magellanic Cloud
(SMC) extinction law from \citet{Pei92}, and treating the extinction
as a thin screen.  The SMC law is a reasonable choice since its
metallicity corresponds approximately to the few metallicity
constraints so far available for \lya\ emitters
\citep{Finkelstein11a}.

We next added the directly measured spectroscopic line fluxes
to the model spectrum at the appropriate wavelengths, since these
emission lines are not included in the model output.  This step
follows the application of IGM and dust opacity, since nature has
already applied these effects to the observed line fluxes.

At this point, we have a full model spectrum accounting for stellar
populations, emission lines, dust, and intergalactic hydrogen
absorption.  We multiply this spectrum by the bandpass of each 
filter in the photometric data set, integrate, and normalize
appropriately to obtain model fluxes in each observed filter for the
model under consideration.  These can be compared to the observed data
to obtain a goodness-of-fit parameter $\chi^2 = \sum_{j=1}^N
(f_{j,obs} - f_{j,mod})^2 / (\delta f_{j,obs})^2$. 

To optimize the model, we fitted the observed spectral energy
distribution of the sample galaxies by a simple Monte Carlo approach
that randomly varies the amount of dust and the mass in stars in each
of 13 logarithmically spaced age bins, and accepts a change to the
model parameters using $\chi^2$ minimization.  This approach allows a
more general star formation history than a single burst or an
exponentially decaying star formation rate.  While the resulting
sampling of stellar ages is somewhat coarse compared to a typical
single burst model, the associated uncertainty is not a dominant
factor in our stellar mass estimates.  The model fits usually converge
to a case where 1--3 of the 13 mass bins dominate the luminosity at
all observed wavebands.  We therefore ``trim'' the parameter list by
fixing the stellar mass to zero in bins that are clearly of minor
importance, and rerun the fit with only the important mass bins.  This
``trimming'' step generally has a negligible effect on the final $\chi^2$,
confirming that 1--3 simple stellar populations can explain the
observed spectrum as well as a more complex star formation history.

To further explore the parameter space of acceptable fits, we modified
our code to optimize for either minima or maxima of either stellar
mass or dust extinction, subject to the constraint that the model
$\chi^2$ remain close to the $\chi^2$ of the best fitting model
for each object.  We explored models with
$\Delta \chi^2 = +1$ and $\Delta \chi^2 = +4$.  This corresponds
approximately to the $1\sigma$ and $2\sigma$ error regions in the
parameter space.

\section{The Stellar Mass Tully-Fisher Relation}
\label{sec:tf}
The stellar mass Tully-Fisher (SMTF) relation is a correlation between
the kinematic line widths of galaxies and their stellar masses.  The
SMTF relation is more robust to differences in stellar population
mass-to-light ratio than the original Tully-Fisher relation (which
correlates luminosity with line width; \citet{Tully77}).  The relation
is further generalized by \citet{Kassin07}, who demonstrated that
replacing the circular speed $V_{c}$ with the kinematic estimator $S_{0.5} = 
\left(0.5 V_{rot}^2 + \sigma^2\right)^{1/2}$ results in an SMTF that is
both tighter and more applicable to the wide range of galaxy properties
seen at high redshift.   $S_{0.5}$ has an additional advantage: 
For spatially unresolved galaxies (like those we study here),
$S_{0.5}$ can be measured reasonably accurately regardless
of whether the width is dominated by ordered rotation or by random motions.

Our actual measurement is a single number, the line width,
characterized by the observed line-of-sight velocity dispersion $\sigma_{obs}$.
We then use $S_{0.5} = \sigma_{obs}$ for our galaxies.
Consider pure circular motion with a flat rotation curve of velocity $V_c$, 
viewed edge-on: we will find $\sigma_{obs}^2 = \langle (\sin(\phi) V_c)^2 
\rangle= 0.5 V_{c}^2$, so that $S_{0.5} = \sigma_{obs}$.  If instead 
the motion is entirely random, with $V_{rot} = 0$, we expect
$\sigma_{obs} = \sigma$, and again, $\sigma_{obs} = S_{0.5}$. 
Only for ordered rotation in a face-on configuration do we expect
$\sigma_{obs}$ to be a significant under-estimate of $S_{0.5}$.
While we cannot rule out this possiblity with our data, 
it is likely that these galaxies are dynamically hot
($\sigma / V_{rot} \not\ll 1$), since the stellar 
populations dominating the observed light 
are at most $\sim 1$ dynamical time old.

\begin{figure}
\plotone{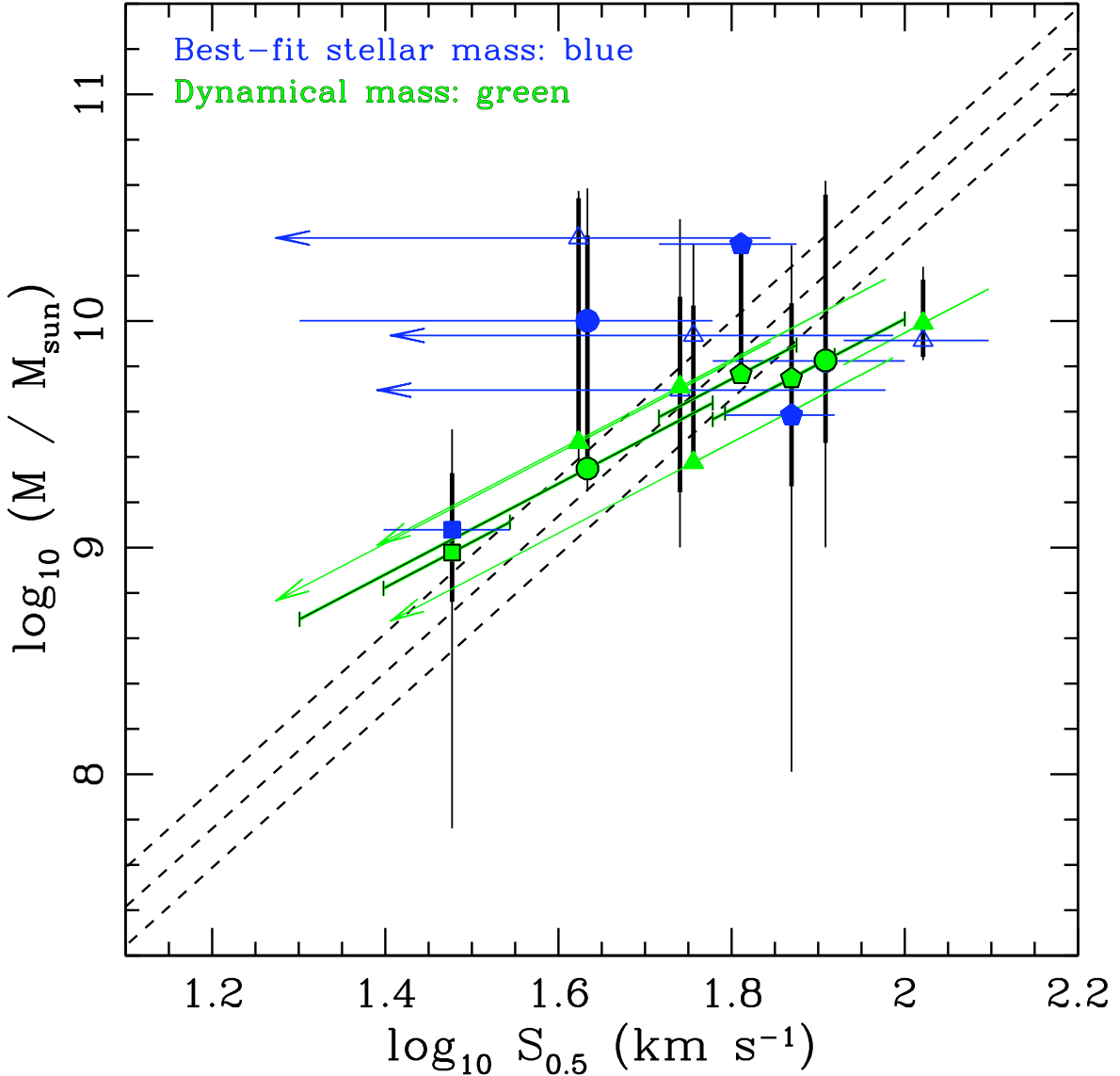}
%% equivalent to  \plotone{smtf_fig_cv8.ps}
% \includegraphics[scale=0.6]{smtf_fig_cv8.pdf}
\caption{Stellar and dynamical mass Tully-Fisher relation.
Colored points mark the \lya\ emitting galaxies
for which we present dynamical mass estimates.  For each, we plot
the best fitting stellar mass (blue) and the dynamical mass (green).
The vertical bar through each point-pair spans the range of stellar
masses for acceptable models (thicker line: models with $\Delta \chi^2 <1$;
thinner line: models with $1< \Delta \chi^2 < 4$).
The error bars on dyamical masses are diagonal, since uncertainty
in velocity dispersion affects both plotted quantities.  
Point shapes correspond 
to surveys:  Circles $\leftrightarrow$ Bok telescope $z=3.1$ survey;
pentagons $\leftrightarrow$ HETDEX pilot survey;
triangles $\leftrightarrow$ \citet{Hashimoto12} Subaru survey; and
square $\leftrightarrow$ \citet{Nilsson11} survey.
Objects with marginally resolved lines are plotted using their
best-estimate values of $\sigma$, but can be regarded as upper
limits on both $S_{0.5}$ and $\mdyn$.
Dashed lines are the best-fit stellar mass Tully-Fisher relation
from \citet{Kassin07}.
}
\label{fig:smtf}
\end{figure}

Our results are shown in figure~\ref{fig:smtf}, both for stellar masses
from SED fitting and for dynamical masses from line width and spatial extent.
The error bars for the stellar masses show the ranges of stellar mass
permitted by models with $\Delta \chi^2 \equiv \chi^2 - \chi^2_{min} < 4$,
and with $\Delta \chi^2 \le 1$.  The error bars for dynamical mass
are determined by the uncertainties in $\sigma$, and are diagonal
since the x-axis of the plot {\it is} $\sigma$.  For three objects
from the \citet{Hashimoto12} sample, $\sigma$ and $\mdyn$ are plotted
as upper bounds.

Comparing our \lya\ sample to the stellar mass Tully-Fisher
relation that \citet{Kassin07} reported for $0.1<z<1.2$ star-forming galaxies,
we find that both the stellar masses
and dynamical masses of LAEs generally follow the established relation.
Dynamical mass estimates show less scatter than stellar masses,
as might be expected given the vagaries of star formation
histories.  

About half the galaxies are consistent with stellar masses falling a
factor of two or more below their dynamical masses, when we account
for the range of acceptable estimates for both.  
For these galaxies, it is possible that the dynamical mass within the
central 1--2 kpc is dominated not by the young stars that power the
observed \lya\ emission, but by some other unseen component.
Collisionless dark matter should not be so strongly concentrated in
the central kpc of a dark matter halo.  Old stars (formed from
centrally concentrated gas $\gg 10^8$ years ago) are a possible
alternative.  The dynamical mass estimates we present are in fact a
tighter limit on the total mass in old stars within the inner $\sim
\kpc$ than are the photometric limits. 
Given the actively star-forming nature of these \lya\ emitting
galaxies, a reservoir of gas is the most intriguing possibility
for the ``excess'' dynamical mass.  Overall, though, such 
excess mass is merely permitted and not required by the data-- so,
overall, the \lya\ galaxy sample shows consistency with
the stellar mass Tully-Fisher relation for other samples.  Whatever
physical properties allow \lya\ to escape these particular star
forming galaxies, they do not strongly affect the mass-linewidth
relation.

The tight correlation observed between $\mdyn$ and line width
is related to the small and nearly constant physical sizes
of the \lya\ galaxy sample \citep{Malhotra12}.  A fixed 
physical size, combined with variable (and sometimes uncertain)
line widths, can generate the observed slope of the $\mdyn$ - $S_{0.5}$
relation.  The  $M_\star$ - $S_{0.5}$ relation from
\citet{Kassin07} has a somewhat steeper slope.  Presuming that stars form
a fairly large and fairly constant fraction of the dynamical mass, this 
slope could be interpreted as evidence for size-linewidth relation
in the \citet{Kassin07} sample, with $r_e \propto S_{0.5}^\gamma$ for 
$\gamma \sim 1$.

\section{Star formation scaling laws} \label{sec:sfr_scale}
We now take the dynamical mass as an upper bound on the gas mass in these
galaxies, and compare their properties to the scaling relations that 
describe star formation in other galaxy classes.  

We use the gas surface mass density limit $\Sigma_{g,max} = \mdyn / (2 \pi
r_e^2)$.  We can improve this bound by subtracting our stellar mass
estimates.  In practice, our {\it maximum} stellar mass 
always exceeds the dynamical mass, allowing the possiblity that
there is no gas mass.  In most cases, though, our {\it minimum}
stellar masses are below the dynamical masses, and we can
subtract them to yield refined estimates of the gas mass surface
density ($\Sigma_g \la (\mdyn - M_{\star,min})/(2 \pi r_e^2)$). 

The star formation rate and gas surface density can be related
according to scaling laws of the form $\log{\Sigma_{SFR}} \approx \asf
\, {\log \Sigma_g} + \bsf$ (where $\Sigma_{SFR}$ is in $\Msun \yr^{-1}
\kpc^{-2}$, and $\Sigma_g$ is in $\Msun \kpc^{-2}$).  For nearby
spirals, and for distant star-forming BzK galaxies, \citet{Daddi10}
find $\asf = 1.42$ and $\bsf = -9.83$ for our choice of units.  For
submillimeter galaxies and (U)LIRGS, Daddi et al find a parallel but
offset ``starburst'' sequence, with $\bsf \approx -8.93$
(corresponding to $8\times$ more star formation for the same gas
surface density).

We determined $\Sigma_{SFR}$ for our sample using the SFR estimates and
half-light radius measurements discussed in \ref{sec:samp_dat}.  We
have selected a Chabrier IMF for consistency with \citet{Daddi10}.
These values of $\Sigma_{SFR}$ exceed the expectations 
for ``normal'' star forming galaxies by a median factor of 4,
based on $\Sigma_{g,max}$ alone (i.e. assuming that {\it all} the
gravitating mass is gas).  The disagreement is significant at the 
$>3\sigma$ level, relative to the 0.33 dex scatter in the scaling 
relation reported by \citet{Daddi10}.  A factor of 4 would place the
\lya\ galaxies in between the normal and starburst sequences,
though closer to the starburst sequence.
If we use $\Sigma_g = (\mdyn - M_{\star,min})/(2 \pi r_e^2)$
in the scaling relations, we find that the median galaxy in our 
sample is forming stars at {\it twice} the rate expected even under
the starburst scaling. 
The bottom line from this comparison  is that 
\lya\ galaxies likely belong to a family of starbursting
objects that includes ULIRGS and submillimeter galaxies, despite
order-of-magnitude differences in mass and star formation rates.

\begin{figure}
\plottwo{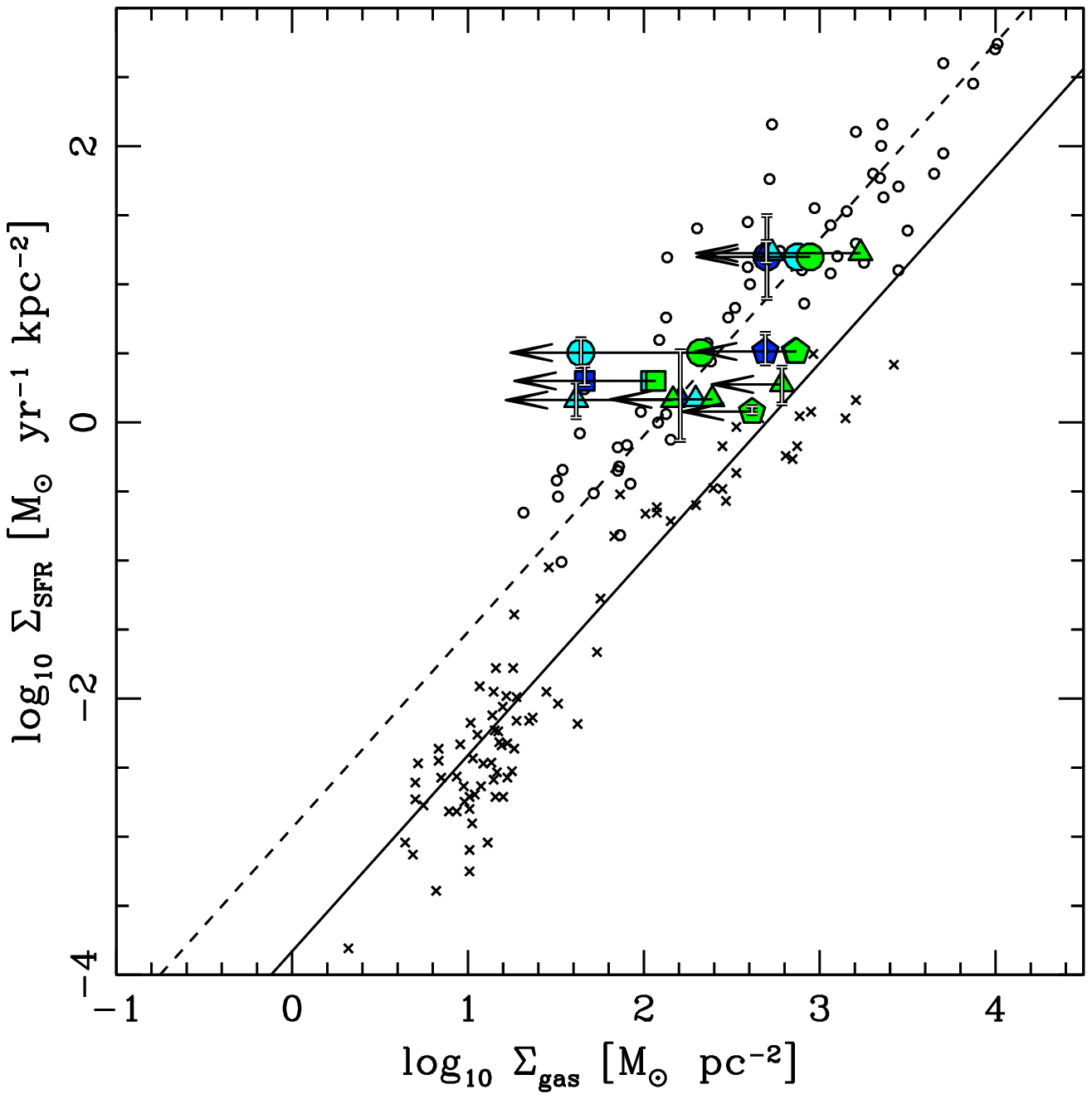}{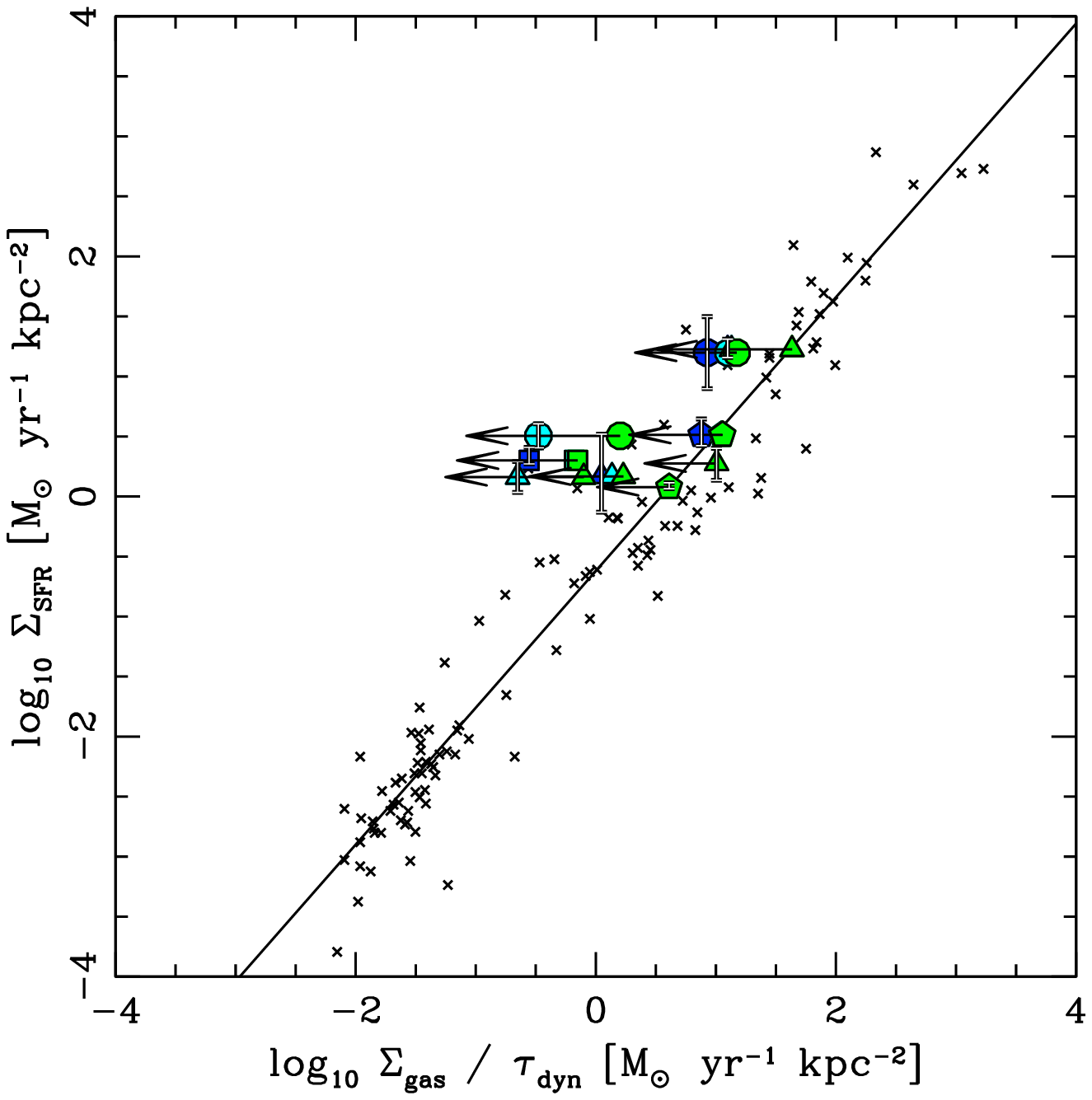}
% \plottwo{sfr_fig1v5.ps}{sfr_fig2v5.ps}
%% \plottwo{sfr_fig1v5.pdf}{sfr_fig2v5.pdf}
%%% \includegraphics[scale=0.6]{sfr_fig1v5.pdf}
%%% \includegraphics[scale=0.6]{sfr_fig2v5.pdf}
\caption{
Star formation law comparison.  {\it First panel:\/}  Relation
between star formation rate surface density $\Sigma_{SFR}$ and 
gas mass surface density $\Sigma_g$, both for our samples (large 
colored points) and for comparison samples drawn from \citet{Daddi10}. 
For each \lya\ galaxy, the green point marks the upper bound on
gas surface density obtained by associating the entire dynamical
mass with gas.  Cyan and blue points take a portion of
the dynamical mass to be associated with stars, using the 
``$2\sigma$'' and ``$1\sigma$'' low-mass models from out stellar 
population fitting.  Error bars in $\Sigma_{SFR}$ account both for
uncertainties in H$\alpha$ flux and in extinction corrections.
Point styles identify subsamples, as in figure~\ref{fig:smtf}.  
The solid line marks the relation for normal star forming
galaxies, and ``x'' points the galaxy samples obeying that relation.
The dashed line marks the starburst
relation, and open black circles mark the corresponding galaxies.  
The \lya\ galaxies are inconsistent with the normal star formation 
relation, and consistent with the starburst galaxy sequence. 
{\it Second panel:\/} $\Sigma_{SFR}$ vs. $\Sigma_g / \tau_{dyn}$.
Since \citet{Daddi10} report a single relation here, all points from
that paper have the same style.  The \lya\ galaxies are consistent with
the general relation, though higher at an insignificant but intriguing
level ($\sim 2 \sigma$).}
\label{fig:sfr_law}
\end{figure}

While the ``normal'' and starburst galaxies follow distinct
$\Sigma_{SFR}$ - $\sigma_g$ relations, they obey a single relation
when $\Sigma_g$ is replaced with the quantity $\Sigma_g / \tau_{dyn}$,
where $\tau_{dyn}$ is the dynamical time.  We use $\tau_{dyn} \approx
2\pi r_e / \sigma$ to place our \lya\ galaxy sample on this relation
also.  While the \lya\ galaxies appear {\it less} unusual when
measured against this relation, they remain systematically above the
trend line found by \citet{Daddi10}.  The difference is suggestive,
rather than significant, being a $2\sigma$ effect.  
The median offset is a factor of $\sim 3$ in  $\Sigma_{SFR}$
at fixed  $\Sigma_g / \tau_{dyn}$.  This is comparable to the $0.44
dex$ scatter in the relation as reported by \citet{Daddi10}).
This possible deviation should be explored using a larger sample.

The range of gas surface densities plotted in figure~\ref{fig:sfr_law}
is from $\sim 30$ to $\sim 1000 \, \Msun \, \hbox{pc}^{-2}$.
Based on a standard ratio of dust to gas column
density, $A_B \approx \Sigma_g / (11 \Msun \, \hbox{pc}^{-2})$
\citep{Bohlin78}, 
this corresponds to $A_B \sim 3$--$90$ magnitudes of extinction.
Yet, our SED fits suggest modest extinctions, $A_v \la 1$, in all cases.  
There are a few possible explanations.
First, the gas surface density could be much lower than
one would expect for the observed level of star formation activity.
In this case, the star formation vs. gas surface density scaling must
be more extreme than even the starburst relation.  Second, the dust-to-gas
ratio could be about 1--2 orders of magnitude lower than in the Milky
Way.  This would be most easily accommodated if the dust-to-gas ratio
scales as the square of metal abundance.  Third, the extinctions inferred
from SED fitting could be dramatic under-estimates.  If so, these objects'
bolometric luminosity would mostly emerge in the rest-frame far
infrared, making them readily detectible with submillimeter imaging
(cf. \citet{Finkelstein09b}).

\section{Densities}\label{sec:density}
Given our estimates of the dynamical mass, it is straightforward
to determine the mean density within the effective radius for
our sample. 
We find
\[ 
\rhoeff = {3\over 2 \pi G } {\sigma^2 \over r_e^2} = {3\pi \over G t_{orb}^2}
\]
where we have used $v_c = \sqrt{2} \sigma$ and $t_{orb} = 2 \pi r_e / v_c$.

\begin{figure}
\plotone{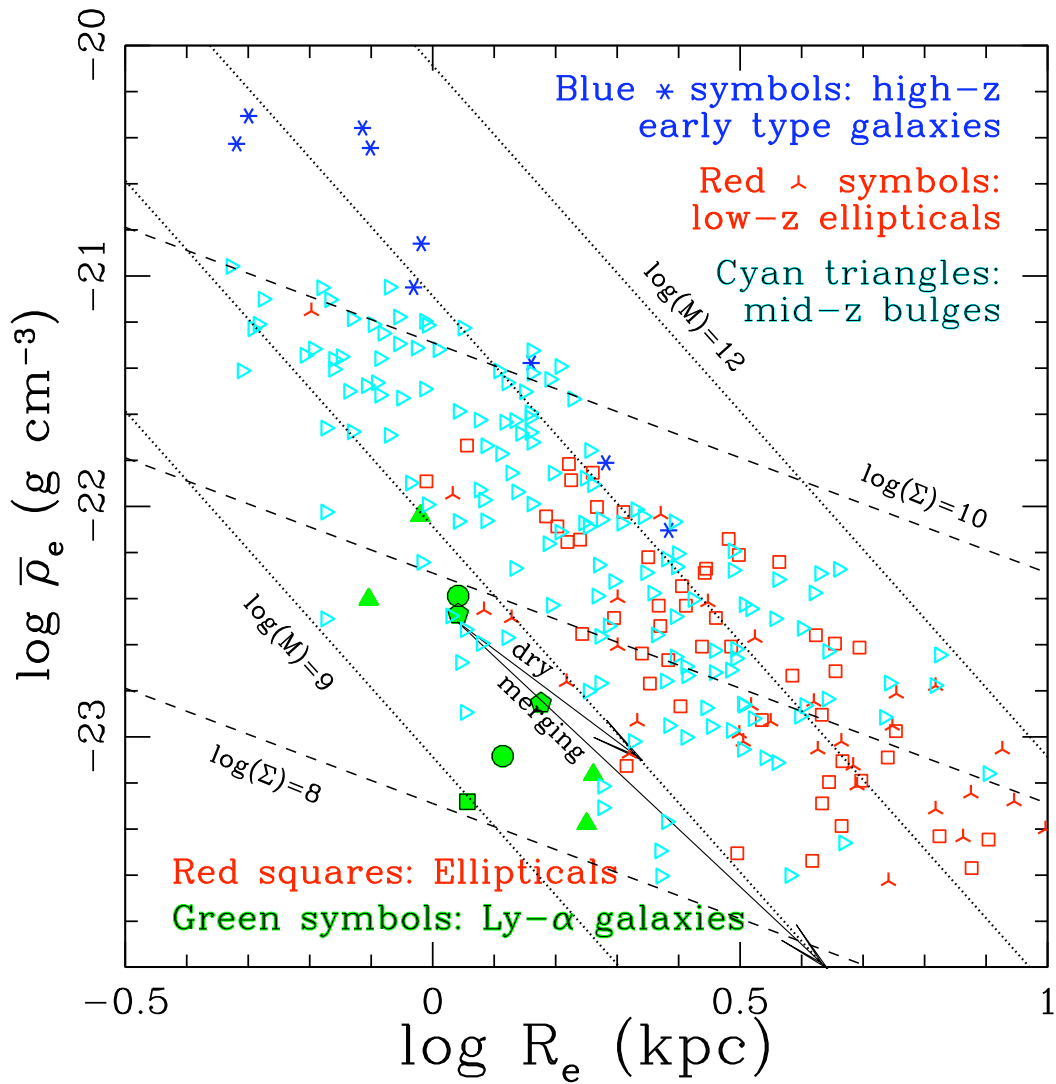}
% equiv. to \plotone{density_fig_v7.ps}
%%\plotone{density_fig_v.pdf}
%%%\includegraphics[scale=0.6]{density_fig_v7.pdf}
\caption{Densities and effective radii for \lya\ galaxies (green
circles) and comparison samples. Cyan triangles are 
intermediate-redshift disk galaxy bulges \citet{MacArthur08}, and red 
squares are low-redshift elliptical galaxies from \citet{Kelson00}, both
with densities derived dynamically.  Blue asterisks are high-redshift 
($z\sim 2.3$) early-type galaxies, and
three-pointed stars are another local early type galaxy sample,
both from \citet{Bezanson09}, and both using stellar masses  estimates 
from SED fitting.
Diagonal dotted
lines mark masses of $10^9$, $10^{10}$, $10^{11}$, and $10^{12}
\Msun$, and dashed lines mark surface densities of $10^8$, $10^9$, and
$10^{10} \Msun\, \kpc^{-2}$.  The \lya\ galaxies are typically about
$10\times$ less massive than the local ellipticals, yet of comparable
density.  Non-dissipative (``dry'') merging results in galaxy
densities that decrease as $\rho \propto 
R_e^{k}$ for $2 \le k \le 2.5$, based on virial theorem
arguments and conservation of total energy (see, e.g.,
\citet{Bezanson09}).  Arrows show such scalings, 
with the shorter arrow corresponding
to one major merger and the longer to a mass doubling through minor mergers.
Neither arrow approaches the region of the elliptical galaxy samples,
indicating that dissipational merging is required if these objects
are indeed progenitors of present day early type galaxies.
\label{fig:dense}}
\end{figure}

Hierarchical structure formation models suggest that most collapsed 
galaxies at $z\sim 3$ should be incorporated into early-type galaxies
or disk galaxy bulges by
the present epoch.  We therefore compare the density
measurements for \lya\ galaxies to corresponding densities
for early-type galaxies and bulges in figure~\ref{fig:dense}.  
The \lya\ galaxy densities are the mean density within the
effective radius, derived dynamically and therefore inclusive of all
gravitating mass. 
For disk galaxy bulges, we take an intermediate-redshift sample 
from \citet{MacArthur08}.  Here we use the published 
effective radii and velocity dispersions to obtain dynamical
estimates of density.
We use two nearby elliptical galaxy samples (one from
\citet{Kelson00}, and one compiled by \citet{Bezanson09} from earlier
work by \citet{Franx89,Peletier90,Jedrzejewski87}).  For the \citet{Kelson00}
sample, we determine densities from published $\sigma$ and $r_e$
measurements; while for the \citet{Bezanson09} sample, we use plotted
values of $r_e$ and $\bar{\rho}(r_e)$.   The \lya\
galaxies are smaller at fixed density, and less dense at fixed radius,
than are the nearby ellipticals and spiral bulges.  
On the other hand they tend to be
smaller and denser than the handful of local ellipticals that have
similar dynamical masses.
We also compare to a sample of early type galaxies at $z\sim 2.3$
\citep{Bezanson09}, which overlaps our \lya\ sample in redshift.
Here, the \lya\ galaxies are of considerably lower density.  In the
samples from \citet{Bezanson09}, we are using {\it stellar} mass
rather than dynamical mass, since $\sigma$ is generally unavailable
for the high-redshift early type galaxies.  Were dynamical masses
available, we might expect these two samples to shift upwards in
figure~\ref{fig:dense}, but the effect would be modest, since the
central regions of early type galaxies are likely dominated by mass in
stars.

The relation between initial overdensity and epoch of collapse ensures
that bound galaxies at $2 < z \la 3$ will, by the present epoch, be
parts of more massive structures. We expect that the \lya\ galaxies we
observe will grow through some combination of merging and smooth
accretion.  Merging can proceed either without dissipation,
as expected for pure mergers of stellar systems (``dry mergers''); 
or with dissipation, as generally expected for gas-rich objects.  
Conservation of energy and the virial theorem can be combined to 
infer the expected evolution of radius with mass for dry mergers,
both in the limit of equal-mass (``major'') mergers, and in the limit
of large mass ratio (``minor'') mergers.  For major mergers, 
we expect $R_e \propto M$, while for minor, we expect $R_e \propto M^2$
\citep[e.g.,][]{Bezanson09}.  In figure~\ref{fig:dense},
these correspond to $\rho \propto R_e^{-2}$ for major mergers,
and $\rho \propto R_e^{-2.5}$ for minor.  We plot vectors for both
scalings.  These show that the \lya\ galaxies we observe cannot
grow to reproduce the observed masses, sizes, and densities of modern
elliptical galaxies through simple dry merging.  Instead,
dissipational (``wet'') merging is required, in order to grow the
galaxies at approximately constant mass density.

\section{Conclusions}
We have performed the first study of dynamical masses for high
redshift \lya\ galaxies, using a sample of nine objects assembled from
four samples
\citep{McLinden11,Richardson12,Finkelstein11a,Hashimoto12,Nakajima12b,Nilsson11}.
Such studies have been previously impractical, given small sizes that
require HST imaging for size measurements \citep{Malhotra12}, faint
continuum that precludes absorption line measurements, and redshifts
that require near-infrared spectroscopy to access those emission lines
most useful for kinematic line width measurements.
We measure dynamical masses ranging from $10^9\Msun$ to $10^{10}\Msun$.

We combine these dynamical masses with stellar masses, to study the
position of \lya\ galaxies on a version of the stellar mass
Tully-Fisher relation between mass and line width.  To derive stellar
masses consistently for the full sample, we fitted population models
to extensive multiband photometry covering rest wavelengths from \lya\
to beyond the 4000\AA\ break.  This affords constraints on young
stars, older stars, and dust.  
 
We find that the \lya\ galaxies are broadly consistent with the
stellar mass Tully-Fisher relation established at lower redshift.
Thus, whatever physical conditions allow the production and escape
of significant \lya, they do not result in strong departures from
the linewidth-mass relation.  On the other hand, about half the sample
galaxies are consistent with stellar masses significantly below
their dynamical masses.  In these cases, there is a possibility of
a dynamically significant reservoir of gas that is present in the inner
kpc region and provides fuel for ongoing star formation.
The dynamical mass measurements are in fact the most powerful 
constraint we have on the presence of old stars in the sample galaxies.
In most cases, the stellar population fits allow up to $\sim 3\times 10^{10}
\Msun$ of old stars before the resulting light appreciably degrades
the quality of the SED fit.  However, dynamical constraints place a 
tighter limit of $\le 1\times 10^{10} \Msun$ on the total mass
for the galaxies in our sample.

By using the dynamical masses as an upper bound on the gas mass, and using
the H$\alpha$ line measurements and/or stellar population fits to
infer star formation rates, we have examined the way the \lya\
galaxies fall on the scaling relations for star formation.  We
conclude that they form stars more actively than the ``normal'' star
forming galaxy population at comparable gas mass surface density.
Their behavior is consistent with that observed in starburst galaxies,
despite the typically smaller masses and sizes of the \lya\ galaxy
population.

The dynamical masses we infer remain 1--2 orders of magnitude 
below the characteristic dark halo masses inferred from the clustering
of \lya\ galaxies.  Since the ratio of baryonic to dark mass should
be globally uniform at around 14\%, we infer that a significant part of 
the baryonic matter associated with these halos has not yet
accreted to the central \kpc, where the active star formation is
observed.

Finally, we examined densities of these objects.  While hierarchical 
structure formation models suggest that small galaxies at $z\sim 2$--3 
should become parts of early type galaxies or spiral bulges by $z=0$,
the \lya\ galaxies are of smaller size and comparable density to
present day elliptical galaxies, and of comparable size but smaller density
than present day spiral galaxy bulges.  If the \lya\ galaxies are to
evolve into either, they must do so through dissipational merging.
This is consistent with the picture of \lya\ galaxies
as young, starbursting galaxies, whose present properties and
future evolution include a large role for gas physics.

\section*{Acknowledgments}
We thank the DARK Cosmology Centre and the Nordea Fonden
in Copenhagen, Denmark, for hospitality during the completion of this work.
We thank Ignacio Ferreras and Sune Toft for helpful discussions.
This work has been supported by the US 
National Science Foundation through NSF grant AST-0808165. 
Based in part on observations obtained at the Gemini Observatory,
which is operated by the Association of Universities for Research in
Astronomy, Inc., under a cooperative agreement with the NSF on behalf
of the Gemini partnership: the National Science Foundation (United
States), the Science and Technology Facilities Council (United
Kingdom), the National Research Council (Canada), CONICYT (Chile), the
Australian Research Council (Australia), Minist\'{e}rio da
Ci\^{e}ncia, Tecnologia e Inova\c{c}\~{a}o (Brazil) and Ministerio de
Ciencia, Tecnolog\'{i}a e Innovaci\'{o}n Productiva (Argentina).
This work was supported in part by a NASA Keck PI Data Award, administered by
the NASA Exoplanet Science Institute. Some data presented herein were
obtained at the W. M. Keck Observatory from telescope time allocated
to the National Aeronautics and Space Administration through the
agency's scientific partnership with the California Institute of
Technology and the University of California. The Observatory was made
possible by the generous financial support of the W. M. Keck
Foundation.
The authors wish to recognize and acknowledge the very significant
cultural role and reverence that the summit of Mauna Kea has always
had within the indigenous Hawaiian community. We are most fortunate to
have the opportunity to conduct observations from this mountain.

\begin{table}
\centering
\begin{minipage}{\linewidth}
\caption{Basic observational properties of the sample.}
\label{tab:1}
\begin{tabular}{llllll}
ID & z & $\sigma$ & $r_e (\hbox{kpc})$ & SFR & Refs.\footnote{References--- 1. \citet{McLinden11,Richardson12}; 2. \citet{Finkelstein11a,Song13}; 3. \citet{Hashimoto12,Nakajima12b}; 4. \citet{Nilsson11}} \\
\hline
LAE40844  	  & 3.1117   & 81   & 1.1    & 120 & 1 \\
LAE27878  	  & 3.11879 & 43   & 1.3    & 34 & 1 \\
HPS194               & 2.287     & 64.7 & 1.5   & 18 & 2 \\
HPS256      	  & 2.491     & 74.0 & 1.1   & 18 & 2 \\
COSMOS13636 & 2.1621  & 57    & 0.787 & 38 & 3 \\
COSMOS30679 &2.19855 & 55    & 1.825 & 395 & 3 \\
CDFS3865          & 2.173     & 105 & 0.955 & 178 & 3 \\
CDFS6482          & 2.205      & 42  & 1.78    & 71 & 3 \\
LAE-COSMOS-47&2.24654& 30  & 1.139 & 9.2 & 4 \\
\end{tabular}
\end{minipage}
\end{table}

\begin{deluxetable}{lllllllllll}
% \tabletypesize{\small}  \rotate   %% sufficient
\tabletypesize{\scriptsize}
\tablewidth{0pt} % set "natural" width for table
\tablecaption{Best fitting SED model for each object.
\label{tab:bestfit}
}
% \centering
% \begin{minipage}{\linewidth}
% \begin{tabular}{lllllllllll}
\tablehead{
\colhead{ID} & \colhead{Age 1\tablenotemark{a}}
& \colhead{Mass 1}& \colhead{Age 2\tablenotemark{a}} & \colhead{Mass 2}
& \colhead{Age 3\tablenotemark{a}} &
\colhead{Mass 3} & \colhead{M$_{tot}$} & \colhead{A$_B$} & 
\colhead{$\chi^2$} & 
\colhead{N$_{bands}$\tablenotemark{b}} 
}
% \footnote{Number of photometric bands used in SED fitting.  
% Wavelength coverage always extends at least from B band ($0.44 \mu m$) 
%to IRAC Channel 1 ($3.6 \mu m$).} 
% \\
%\hline
\startdata
LAE40844 & 2 & $2.78 \times 10^8$ & 8 & $5.53 \times 10^8$ & 300 & $5.83 \times 10^9$ & $6.67 \times 10^9$ & 0.217 & 42.227 & 11 \\
LAE27878 & 2 & $6.96 \times 10^7$ & 600 & $9.96 \times 10^9$ & 0 & 0 & $1.003 \times 10^{10}$ & 0.164 & 3.269 & 11 \\
HPS194 & 2 & $5.52 \times 10^7$ & 300 & $2.18 \times 10^{10}$ & 0 & 0 & $2.185 \times 10^{10}$ & 0 & 14.92 & 9 \\
HPS256 & 2 & $3.82 \times 10^7$ & 1500 & $3.81 \times 10^9$ & 0 & 0 & $3.84 \times 10^9$ & 0.110 & 2.656 & 9 \\
COSMOS13636 & 4 & $1.06 \times 10^8$ & 300 & $8.53 \times 10^9$ & 0 & 0 & $8.63 \times 10^9$ & 0.258 & 35.15 & 9 \\
COSMOS30679 & 4 & $1.00 \times 10^9$ & 150 & $3.95 \times 10^9$ & 0 & 0 & $4.95 \times 10^9$ & 0.906 & 20.22 & 9 \\
CDFS3865 & 2 & $3.24 \times 10^8$ & 150 & $7.88 \times 10^9$ & 0 & 0 & $8.21 \times 10^9$ & 0.211 & 7.672 & 12 \\
CDFS6482 & 2 & $1.44 \times 10^8$ & 16 & $6.23 \times 10^8$ & 2000 & $2.25 \times 10^{10}$ & $2.325 \times 10^{10}$ & 0.299 & 3.862 & 11 \\
LAE-COSMOS-47 & 4 & $2.63 \times 10^7$ & 300 & $1.03 \times 10^9$ & 0 & 0 & $1.20 \times 10^9$ & 0 & 27.5 & 19% \\
\enddata
\tablenotetext{a}{Ages in Myr.}
\tablenotetext{b}{Number of photometric bands used in SED fitting.  
 Wavelength coverage always extends at least from B band ($0.44 \mu m$) 
to IRAC Channel 1 ($3.6 \mu m$).}
% \end{tabular}
%\caption{Best fitting SED model for each object.
%\label{tab:bestfit}
%}
% \end{minipage}
\end{deluxetable}

\begin{deluxetable}{llllllllll}
\tabletypesize{\scriptsize}
\tablewidth{0pt} % set "natural" width for table
\tablecaption{Range of SED fitting results for the sample. 
\label{tab:sedfits}
}
% \begin{tabular}{llllllllll}
\tablehead{
\colhead{ID}  & \colhead{$M_{-2\sigma}$}  & \colhead{M$_{*,-1\sigma}$}  
& \colhead{M$_{*,+1\sigma}$}  & \colhead{M$_{*,+2\sigma}$}  
& \colhead{$A_{B,-2\sigma}$}  & \colhead{$A_{B,-1\sigma}$}  
& \colhead{$A_{B,best}$} & \colhead{$A_{B,+1\sigma}$} 
& \colhead{$A_{B,+2\sigma}$}
}
% \hline
\startdata
LAE40844    & $1.00\times 10^{9}$ & $2.90\times 10^{9}$ & $3.62\times 10^{10}$ & $4.16\times 10^{10}$ & 0.081 & 0.150 & 0.217 & 0.22 & 0.241 \\
LAE27878    & $1.77\times 10^{9}$ & $4.05\times 10^{9}$ & $2.39\times 10^{10}$ & $3.85\times 10^{10}$ & 0 & 0 & 0.164 & 0.201 & 0.254 \\
HPS194      & $2.05\times 10^{10}$ & $2.12\times 10^{10}$& $2.26\times 10^{10}$& $2.45\times 10^{10}$ & 0 & 0 & 0 & 0.030 & 0.100 \\
HPS256      & $1.03\times 10^{8}$ & $1.87\times 10^{9}$ & $1.20\times 10^{10}$ & $2.16\times 10^{10}$ & 0 & 0.009 & 0.110 & 0.199 & 0.298 \\
COSMOS13636 & $2.61\times 10^{9}$ & $4.49\times 10^{9}$ & $1.17\times 10^{10}$ & $2.20\times 10^{10}$ & 0.083 & 0.161 & 0.258 & 0.291 & 0.320 \\
COSMOS30679 & $1.00\times 10^{9}$ & $1.75\times 10^{9}$ & $1.28\times 10^{10}$ & $2.82\times 10^{10}$ & 0.718 & 0.831 & 0.906 & 0.991 & 1.141 \\
CDFS3865    & $6.72\times 10^{9}$ & $6.94\times 10^{9}$ & $1.52\times 10^{10}$ & $1.74\times 10^{10}$ & 0.184 & 0.196 & 0.211 & 0.230 & 0.259 \\
CDFS6482    & $2.10\times 10^{9}$ & $9.68\times 10^{9}$ & $3.48\times 10^{10}$ & $3.75\times 10^{10}$ & 0.097 & 0.235 & 0.299 & 0.332 & 0.358 \\
LAE-COSMOS-47 & $5.77\times 10^{7}$ & $5.77\times 10^{8}$ & $2.13\times 10^{9}$ & $3.33\times 10^{9}$ & 0 & 0 & 0 & 0.093 & 0.172 \\
% \end{tabular}
\enddata
\end{deluxetable}

\begin{table}
\centering
\label{tab:masses}
\begin{minipage}{\linewidth}
\caption{Masses and densities of the sample.}
\begin{tabular}{lllll}
ID & M$_{dyn}$ & $\log(\bar{\rho}_e)$  & M$_{*,min}$ & M$_{*,max}$ \\
\hline
LAE40844 & $6.7\times 10^{9}$ & -22.39 & $2.90\times 10^9$ & $3.62\times 10^{10}$ \\
LAE27878 & $2.2\times 10^{9}$ & -23.08 & $4.05\times 10^9$ & $2.39\times 10^{10}$ \\
HPS194 & $5.8\times 10^{9}$ & -22.85 & $2.12\times 10^{10}$ & $2.26\times 10^{10}$ \\
HPS256 & $5.6\times 10^{9}$ & -22.47 & $1.87\times 10^9$ & $1.20\times 10^{10}$ \\
COSMOS13636 & $2.4\times 10^{9}$ & -22.40 & $4.49\times 10^9$ & $1.17\times 10^{10}$ \\
COSMOS30679 & $5.1\times 10^{9}$ & -23.16 & $1.75\times 10^9$ & $1.28\times 10^{10}$ \\
CDFS3865 & $9.8\times 10^{9}$ & -22.04 & $6.94\times 10^9$ & $1.52\times 10^{10}$ \\
CDFS6482 & $2.9\times 10^{9}$ & -23.38 & $9.68\times 10^9$ & $3.48\times 10^{10}$ \\
LAE-COSMOS-47 & $9.5\times 10^{8}$ & -23.28 & $5.77\times 10^8$ & $2.13\times 10^9$ \\
\end{tabular}
\end{minipage}
\end{table}

%\bibliographystyle{apj}
%
%\bibliography{lya_refs_v1}

\label{lastpage}
\end{document}